\documentclass[manuscript,screen]{acmart}

\usepackage{makecell}

\AtBeginDocument{%
  \providecommand\BibTeX{{%
    \normalfont B\kern-0.5em{\scshape i\kern-0.25em b}\kern-0.8em\TeX}}}

\copyrightyear{2024}
\acmYear{2024}
\setcopyright{rightsretained}
\acmConference[AutomotiveUI Adjunct '24]{16th International Conference on Automotive User Interfaces and Interactive Vehicular Applications}{September 22--25, 2024}{Stanford, CA, USA}
\acmBooktitle{16th International Conference on Automotive User Interfaces and Interactive Vehicular Applications (AutomotiveUI Adjunct '24), September 22--25, 2024, Stanford, CA, USA}
\acmDOI{10.1145/3641308.3685048}
\acmISBN{979-8-4007-0520-5/24/09}

\newcommand\citetodo[1]{\textcolor{olive}{[CITE]}}

\usepackage{enumitem}
\usepackage{xspace}

\begin{document}



\title[Regaining Trust]{Regaining Trust: Impact of Transparent User Interface Design on Acceptance of Camera-Based In-Car Health Monitoring Systems}





\author{Hauke Sandhaus}
\email{hgs52@cornell.edu}
\orcid{0000-0002-4169-0197}
\affiliation{%
  \institution{Cornell University, Cornell Tech}
  \streetaddress{2 West Loop Rd}
  \city{New York}
  \state{New York}
  \country{USA}
  \postcode{10044}
}
\author{Madiha Zahrah Choksi}
\orcid{0009-0008-4752-7164}
\email{mc2376@cornell.edu}
\affiliation{%
  \institution{Cornell Tech}
  \streetaddress{2 West Loop Rd}
  \city{New York}
  \state{New York}
  \country{USA}
  \postcode{10044}
}

\author{Wendy Ju}
\orcid{0000-0002-3119-611X}
\email{wendyju@cornell.edu}
\affiliation{%
  \institution{Cornell Tech}
  \streetaddress{2 West Loop Rd}
  \city{New York}
  \state{New York}
  \country{USA}
  \postcode{10044}
}


\renewcommand{\shortauthors}{Sandhaus}


\begin{abstract}
Introducing in-car health monitoring systems offers substantial potential to improve driver safety. However, camera-based sensing technologies introduce significant privacy concerns. This study investigates the impact of transparent user interface design on user acceptance of these systems. We conducted an online study with 42 participants using prototypes varying in transparency, choice, and deception levels. The prototypes included three onboarding designs: (1) a traditional Terms and Conditions text, (2) a Business Nudge design that subtly encouraged users to accept default data-sharing options, and (3) a Transparent Walk-Through that provided clear, step-by-step explanations of data use and privacy policies. Our findings indicate that transparent design significantly affects user experience measures, including perceived creepiness, trust in data use, and trustworthiness of content. Transparent onboarding processes enhanced user experience and trust without significantly increasing onboarding time. These findings offer practical guidance for designing user-friendly and privacy-respecting in-car health monitoring systems.
\end{abstract}

%

\ccsdesc[500]{Security and privacy~Usability in security and privacy}
\ccsdesc[500]{Human-centered computing~Usability testing}
\ccsdesc[100]{Human-centered computing~Empirical studies in HCI}
\ccsdesc[500]{Human-centered computing~Interface design prototyping}

 \keywords{In-Car Health Monitoring; Privacy; User Experience; Transparent Design; Creepiness; Trustworthiness; Pervasive Sensing; Technology Acceptance; Online Study; Vehicle}




\maketitle

\section{Introduction}
In recent years, there is growing interest in integrating health monitoring systems into cars to provide real-time monitoring and support for individuals with chronic health conditions~\cite{Transparency-Market-Research2023-dy}. These systems commonly utilize camera-based technologies to capture vital health information to enhance driver safety by detecting health emergencies in real-time and alerting emergency services. The potential benefits of such systems are substantial, including improved response times to medical emergencies and better overall health outcomes for drivers. In the European Union, driver monitoring systems will be mandated by law ~\cite{Gospodinova2022-wh}. 

However, the use of camera-based sensing technologies in such systems raises significant privacy concerns~\cite{Chen2018-ni}. Users may be apprehensive about continuous monitoring and the potential misuse of their personal health data in the sensitive automotive context with often weak privacy protections~\cite{Caltrider2023-jr, Bollinger2017-ad}. Ensuring user acceptance of these technologies requires addressing these privacy concerns through thoughtful design choices.

This study investigates the impact of transparent user interface design on users' experience and acceptance of camera-based health monitoring systems. We address two primary research questions: (1) How does transparent user interface design impact users' experience of a health monitoring system? (2) Do users with health conditions have aversions to camera-based health monitoring systems in cars?

This paper outlines our methodology for investigating these questions, including study design, participant recruitment, data collection, prototype design, user perception assessment, data analysis, and statistical tests. The study aims to inform the design of user-friendly, privacy-respecting in-car health monitoring systems.

\section{Related Work}
The adoption of new technologies, particularly those that are potentially intrusive, often faces significant privacy concerns. In-car camera-based health monitoring systems can raise substantial privacy issues due to their ability to capture detailed visual information~\cite{VisualPrivacyProtectionSurvey}. Prior work suggests that strategies such as transparent user interface design and appropriate onboarding could be effective in addressing these concerns, making the system's operations and data usage clear to users, which can help alleviate privacy concerns and increase acceptance \cite{EndUserPrivacyHCI, LongSurveillance}.

Transparency in design is crucial for building trust between users and technology providers. Research has shown that when users understand how their data is collected, used, and protected, they are more likely to trust and accept new technologies \cite{Esmaeilzadeh2019-mh}. For example, providing clear explanations of data usage policies and offering users control over their data can significantly enhance trust and acceptance \cite{Wanner2022-qe}.

However, researchers also discuss the transparency paradox, where detailed privacy policies are less likely to be understood if they state all the possible conditions for the use of personal data. The concept of the transparency paradox, also known as the second privacy paradox, highlights the complexity of providing users with sufficient information to make informed privacy decisions without overwhelming them \cite{Hilden2016-qo}. Effective transparency requires balancing detailed disclosures with user comprehension \cite{Bruening2015-rw}. Notices are often found to be complex, unclear, and too lengthy to be useful to consumers or to support meaningful choices; here, more transparency often results in adverse effects. The need for better privacy policies and contracts has become an issue of design~\cite{Huovinen2021-ug}.

Other work has demonstrated that users desire privacy explanations \cite{Brunotte2022-vb}, but more transparency through explanations in advice-giving systems does not always lead to more trust, indicating there is an appropriate level of transparency that relates to user understanding \cite{Zhao2019-hi}. Moreover,  \citet{Knowles2023-tt} explore the relationship between privacy and trust in apparent privacy paradox circumstances, highlighting the importance of hopeful trust in moderating people's willingness to disclose personal information.

In response to privacy concerns, alternative sensors such as radar are actively being developed. These sensors can detect movement and gestures without capturing detailed visual information, potentially making them more acceptable to users \cite{Lien2016-qt}. Efforts have also been made to improve the sensing capabilities of these alternative sensors. For instance, \citeauthor{Vid2Doppler} used video data to train machine learning algorithms for mmWave sensing, aiming to offer similar activity tracking as cameras but without the privacy implications \cite{Vid2Doppler}. A recent review by \citeauthor{Stewart2024-uy}, however, indicates that less 'obtrusive' on-device AI sensors might pose worse privacy concerns \cite{Stewart2024-uy}. 

The concept of ``creepiness'' in human-computer interaction refers to the unsettling feeling that users may experience when interacting with a technology \cite{Langer2018-po, Seberger2022-pt}. This feeling can be exacerbated by a lack of transparency and control over personal data. Addressing creepiness through transparent design and user control mechanisms is essential for overcoming initial resistance to new technologies \cite{Tene2013-nj}.

Research also highlights the lack of adequate camera monitoring feedback and proposes various new design strategies to address this issue \cite{Koelle2018-ey}. For instance, \citeauthor{Egelman2015-am} suggest crowdsourcing privacy indicators for ubiquitous sensing platforms, which can enhance user understanding and trust. Additionally, various visual privacy protection approaches, including pixelation, blurring, and avatar replacement, have been applied to reduce perceived invasiveness \cite{FiltersAwarnessPrivacy, VisualPrivacyProtectionSurvey}.

Within the AutoUI community, the need for alternatives to privacy-concerning in-vehicle data collection is recognized. Recent studies have shown limited investigation for improved privacy in driver state monitoring. For example, \citeauthor{De_Salis2019-rr} explored noninvasive distraction detection methods that do not rely on psychophysiological data, and \citeauthor{Ihme2021-kq} developed and evaluated a data privacy concept for voice-based driver state monitoring \cite{De_Salis2019-rr, Ihme2021-kq}.

\section{Methodology}
We conducted an online study with 42 participants to explore the impact of transparent user interface design on user acceptance of camera-based in-car health monitoring systems. Participants interacted with prototypes that varied in transparency, choice, and deception levels.

A total of 42 participants were recruited using Prolific\footnote{\url{https://www.prolific.co}}, with a focus on achieving a balanced and diverse sample from the U.S., following common guidelines for sample size~\cite{Moran2021-ik}. Participants were car users with long-term health conditions or disabilities. The participants' ages ranged from 20 to 75 years, with an average age of approximately 41 years and a standard deviation of 14.84. The gender distribution was balanced, with 19 females and 21 males. The majority of the participants were White (33 out of 40). All participants were U.S. citizens, although 4 were born outside the U.S. Two participants did not provide demographic information. 

Participants engaged in an online experiment where they interacted with two prototypes in six conditions hosted online. After interacting and observing each prototype condition, participants completed a series of questionnaires measuring perceived user experience using 7-point Likert scales. The order of prototype conditions was randomized. Participants were paid \$2.60, which amounts to \$11.64/hr on average for the study, which took a median time of 13:24 minutes. The time spent assessing each prototype was measured.

The study focused on the impacts of onboarding, including terms of use and privacy policy, and live view, with real-time monitoring feedback. Mockups were created with Apple Freeform and translated into semi-functional online prototypes using Framer \cite{Framer_2022}. Onboarding required users to accept or decline the terms of use and decide whether to share data with the technology provider. Three designs were used: (1) Terms of Use (T\&C): Typical legal text-based user onboarding. (2) Business Nudge: Deceptive designs to nudge users to accept default business-preferred options \cite{Acquisti2017-sa}. (3) Transparent Choice: Multi-step onboarding process explaining in chunks what the user agrees to and the options they have \cite{Egelman2015-am}.

The live view prototype conditions differed in live camera feedback: (1) No Camera: No live camera feedback. (2) Camera View: Live camera feed of the participant, overlaid with a mesh of facial landmarks. (3) Blurred Camera: Blurred live camera feed of the participant, overlaid with a mesh of facial landmarks.

To contribute to open science practices in AutoUI research~\cite{Ebel2023-sa}, the IRB documentation, survey measures, stimuli (prototype videos), and data are available in the OSF repository\footnote{\url{https://osf.io/yp6bs/}}. 

\begin{figure*}[t]
      \centering
       \includegraphics[trim=0cm 0cm 0cm 14cm, clip=true, width=0.6\textwidth]{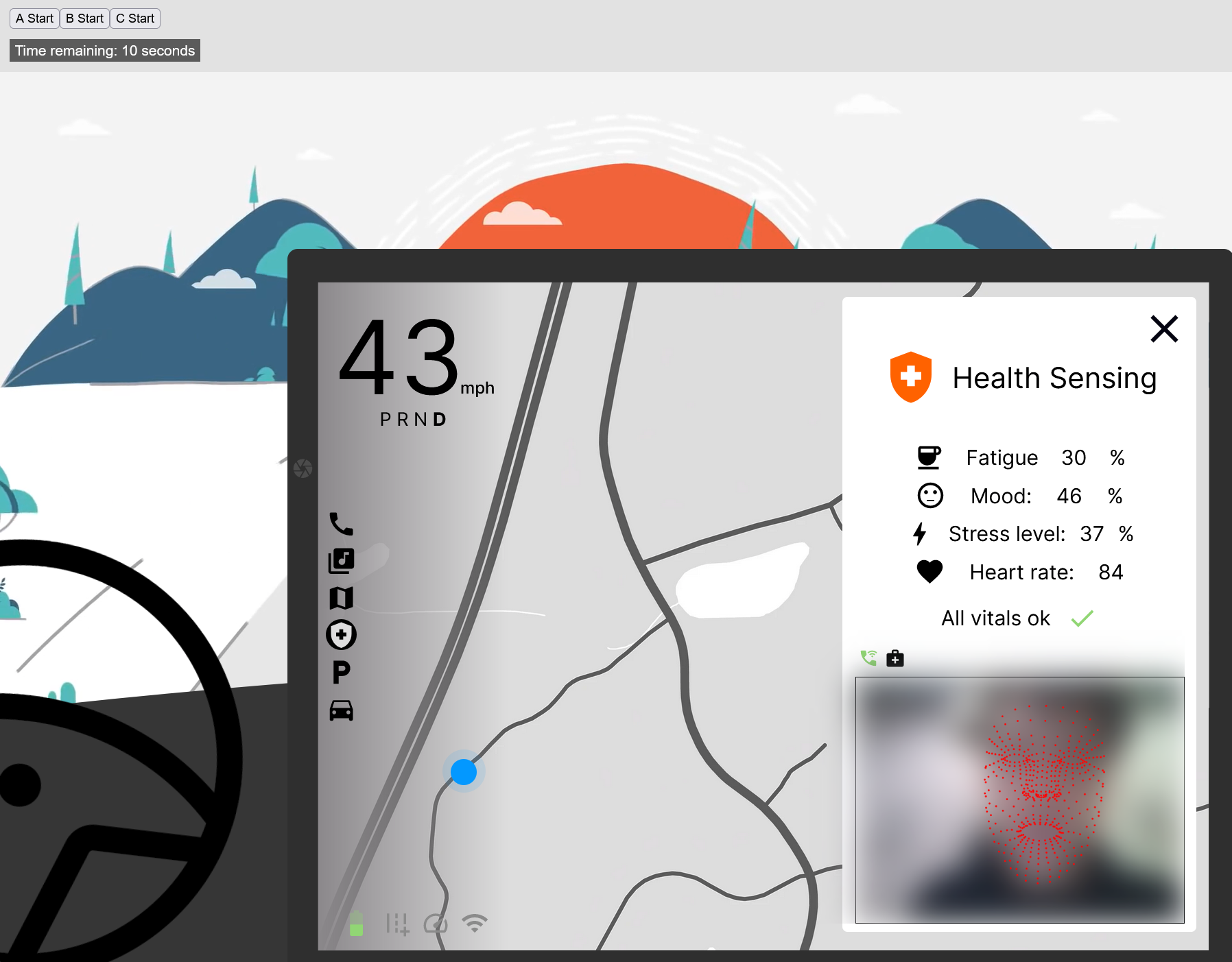}
\captionsetup{width=\textwidth}   
\caption{Tested Live View Prototype with Blurred Camera. Available at \url{https://health-sensing-car.framer.website}}
      \Description{Tested Live View Prototype with Blurred Camera. Available at \url{https://health-sensing-car.framer.website}}
    \label{fig:proto-live}
\end{figure*}

\begin{figure*}[t]
      \centering
       \includegraphics[trim=26cm 0cm 0cm 13cm,, clip=true,height=0.3\textwidth]{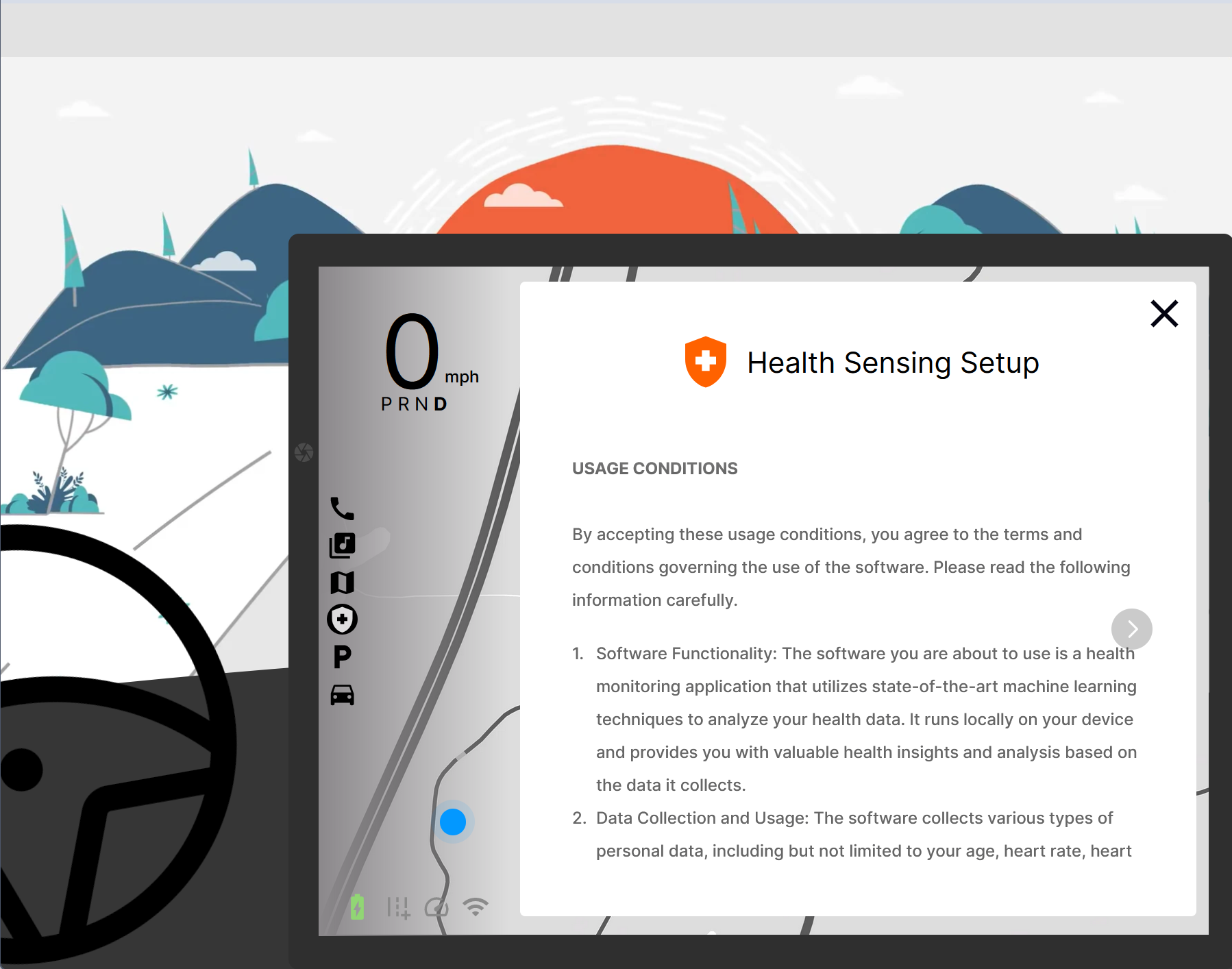}
      \includegraphics[trim=26cm 0cm 0cm 13cm, clip=true,height=0.3\textwidth]{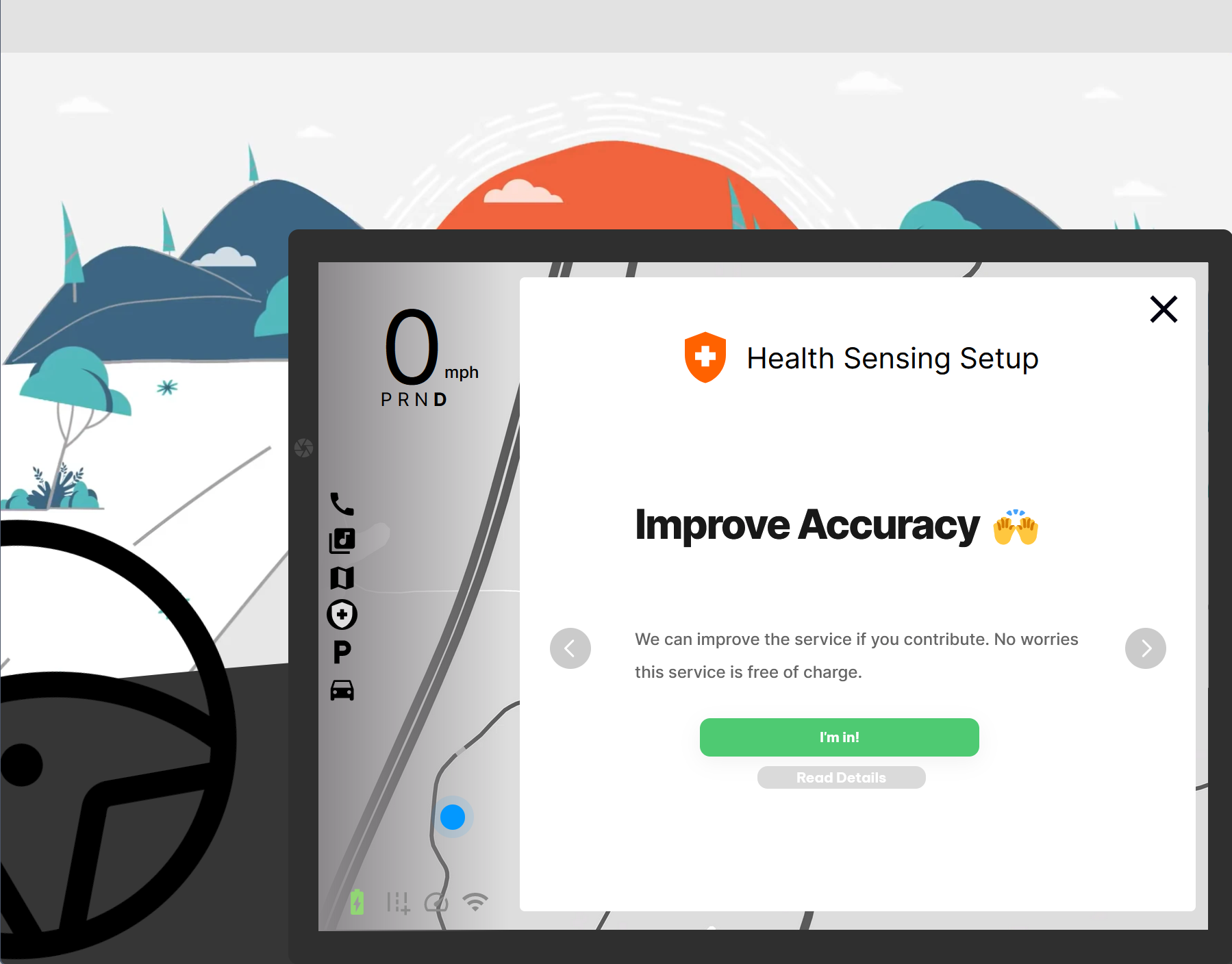}
    \includegraphics[trim=26cm 0cm 0cm 13cm, clip=true,height=0.3\textwidth]{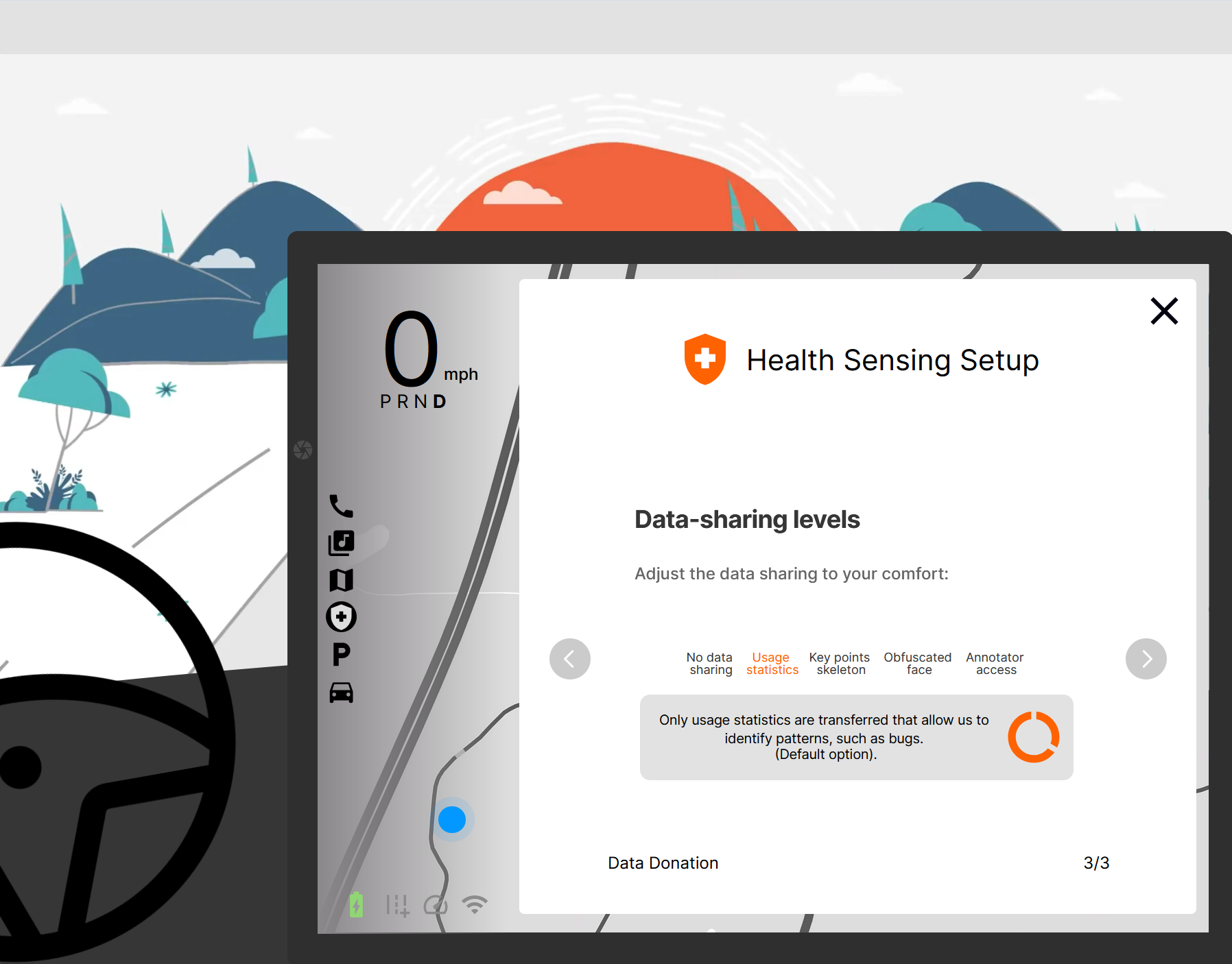}

\captionsetup{width=\textwidth}   
\caption{Tested Onboarding Prototypes. Left to right: Terms and Conditions, Opt-in Nudge, Transparent Walk-through. Available at \url{https://health-sensing-car.framer.website/X}, \href{https://health-sensing-car.framer.website/Y}{/Y}, and \href{https://health-sensing-car.framer.website/Z}{/Z}.}
      \Description{Tested Onboarding Prototypes. Left to right: Terms and Conditions, Opt-in Nudge, Transparent Walk-through}
    \label{fig:proto-onboarding}
\end{figure*}

Data analysis was conducted using R Studio. For the statistical analysis, we first inspected the normality and homogeneity of variance assumptions using Shapiro-Wilk and Levene's tests, respectively. Depending on the results of these tests, either parametric or non-parametric tests were conducted to analyze the data. Descriptive statistics, including mean scores, standard deviations, p values, and effect sizes, were computed for each scale. Violin plots were created with the ggplot2 and ggstatsplot packages for R.

The chosen measures included:
\begin{itemize}
    \item \textit{Pragmatic Quality}: Assessed the perceived usability of the technology and its user interface, based on the User Experience Questionnaire (UEQ) \cite{Schrepp2017-tl}.
    \item \textit{Hedonic Quality}: Evaluated the stimulation and novelty of the technology and its user interface, also derived from the UEQ \cite{Schrepp2017-tl}.
    \item \textit{Creepiness}: Measured the degree to which participants found the technology unsettling, inspired by previous work on privacy and human-computer interaction \cite{Langer2018-po, Seberger2022-pt}.
    \item \textit{Trustworthy Content}: Assessed the perceived usefulness and trustworthiness of the information provided by the product, based on the modular UEQ extension \cite{Schrepp2019-dq}.
    \item \textit{Data Use (Trust)}: Measured the perceived security and reliability of the product's use of personal data, using scales from the UEQ extension \cite{Hinderks2016-tv}.
\end{itemize}

\section{Findings}
Our findings indicate that participants perceived the Transparent Walk-Through onboarding condition as the least creepy (mean = -1.37), and the Blurred Camera view was perceived as less creepy (mean = -0.63) than the Camera View (mean = -0.13). The Friedman test for creepiness scores showed significant differences across conditions for both onboarding, $\chi^2(2) = 21.64, p < 0.001$, and live view prototypes, $\chi^2(2) = 8.33, p = 0.02$.

Participants reported the highest user experience with the Transparent Walk-Through onboarding procedure, with mean pragmatic quality scores of 1.57 and mean hedonic quality scores of 1.53. The Friedman test indicated significant differences in user experience scores between onboarding conditions for both pragmatic quality, $\chi^2(2) = 9.23, p = 0.009$, and hedonic quality, $\chi^2(2) = 7.93, p = 0.019$.

Trust in data use was significantly higher with the Transparent Walk-Through onboarding procedure (mean = 1.08) compared to the Opt-in Nudge design (mean = 0.30). The Friedman test showed significant differences in trust scores across onboarding conditions, $\chi^2(2) = 8.14, p = 0.02$. However, there were no significant differences in trust scores between the live view camera feedback conditions, $F(1.56, 63.8) = 1.74, p = 0.19$.

Trustworthiness of content was also significantly higher with the Transparent Walk-Through onboarding procedure (mean = 1.36) compared to the Opt-in Nudge design (mean = 0.67). The Friedman test showed significant differences in trustworthiness scores across onboarding conditions, $\chi^2(2) = 6.48, p = 0.04$. There were no significant differences in trustworthiness scores between the live view camera feedback conditions, $\chi^2(2) = 1.49, p = 0.48$.

There were no significant differences in hedonic quality between the live view camera feedback conditions, although the means for the Blurred Camera (mean = 5.1) and Camera View conditions (mean = 5.0) were marginally higher than the No Camera condition (mean = 4.8), $F(2, 39) = 1.45, p = 0.25$.

Engagement time, measured as the last click interacting with the prototype, before evaluating it, showed significant differences across onboarding conditions, with the Transparent Walk-Through having a mean time of 109.01 seconds, Terms and Conditions having 103.27 seconds, and Business Nudge having 87.73 seconds, $\chi^2(2) = 9.22, p = 0.01$. However, participants did not spend significantly more time on the terms of use condition compared to the other two conditions, $F(2, 39) = 0.98, p = 0.38$, suggesting that the transparent walk-through process does not lead to significantly longer engagement times.

\begin{figure*}[t]
      \centering
       \includegraphics[width=0.45\textwidth]{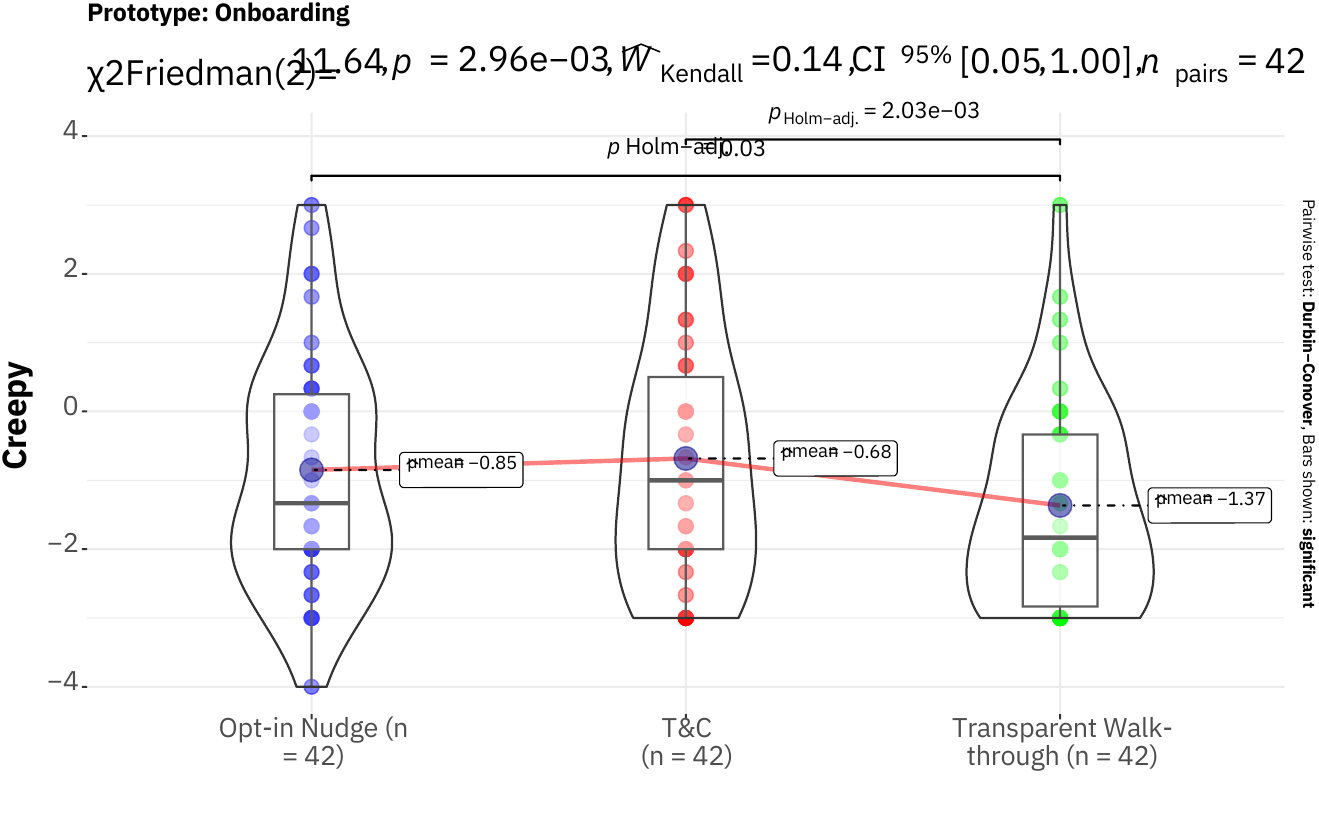}
       \includegraphics[width=0.45\textwidth]{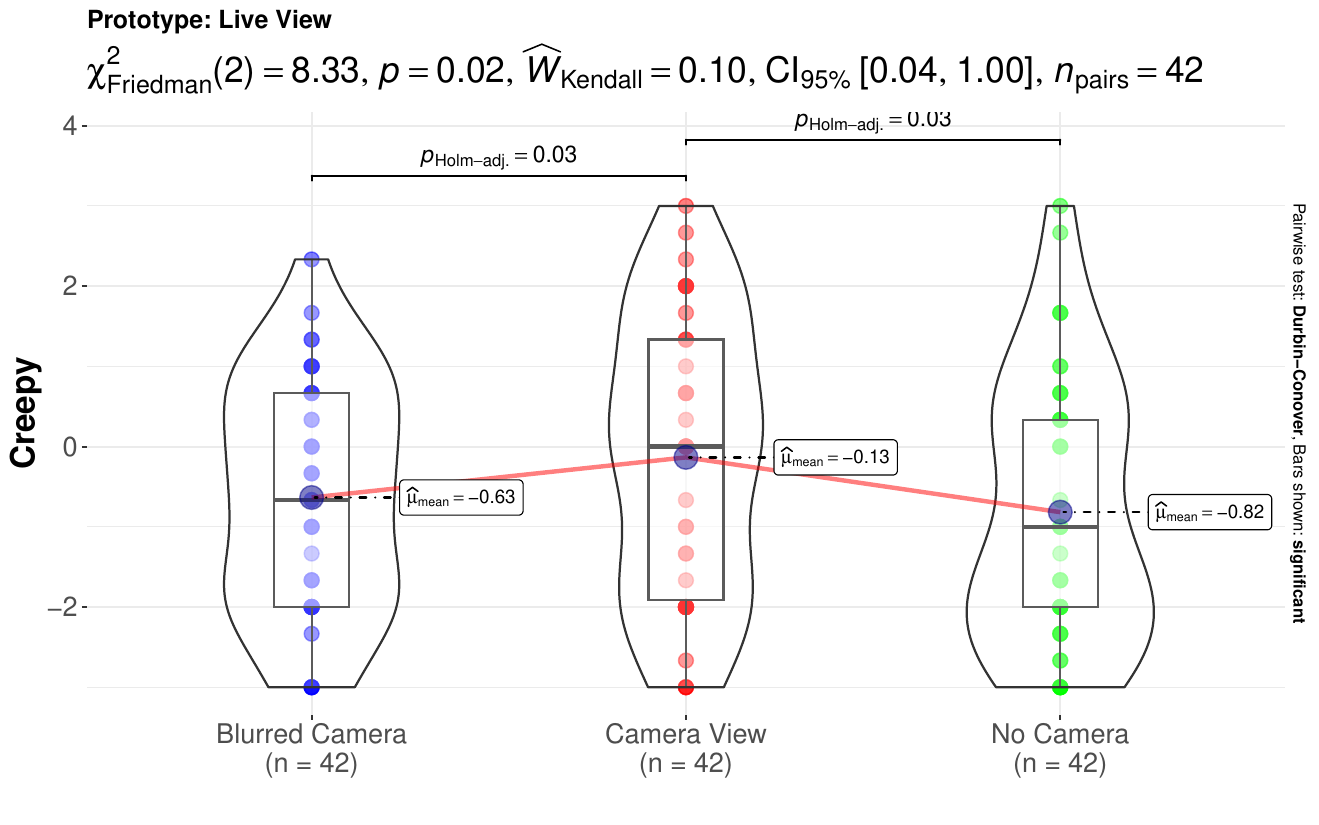}
\captionsetup{width=0.6\textwidth}   
\caption{Perceived Creepiness of the Onboarding and Live View Prototypes.}
      \Description{ Perceived Creepiness of the Onboarding and Live View Prototypes.}
    \label{fig:Creep}
\end{figure*}

\begin{figure*}[t]
      \centering
    \includegraphics[width=0.45\textwidth]{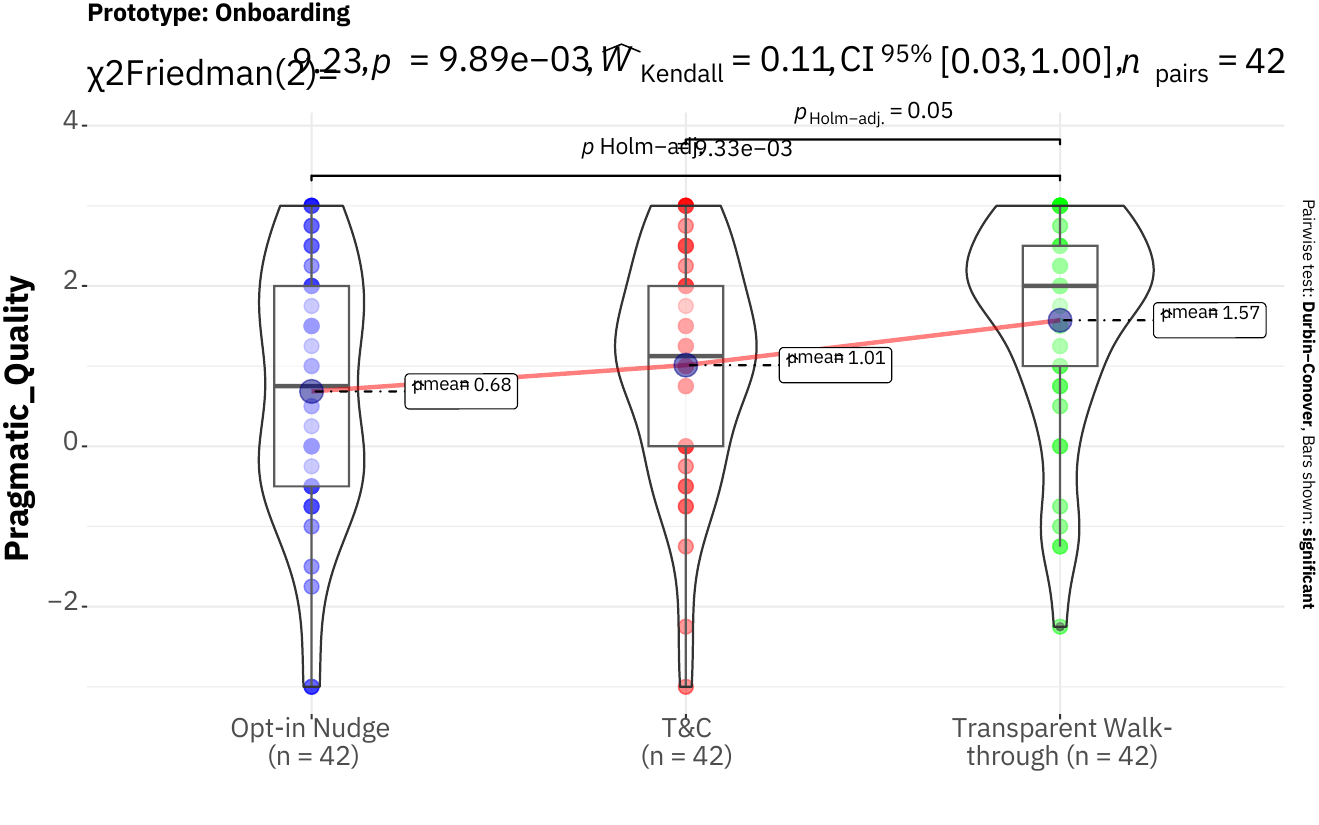}
    \includegraphics[width=0.45\textwidth]{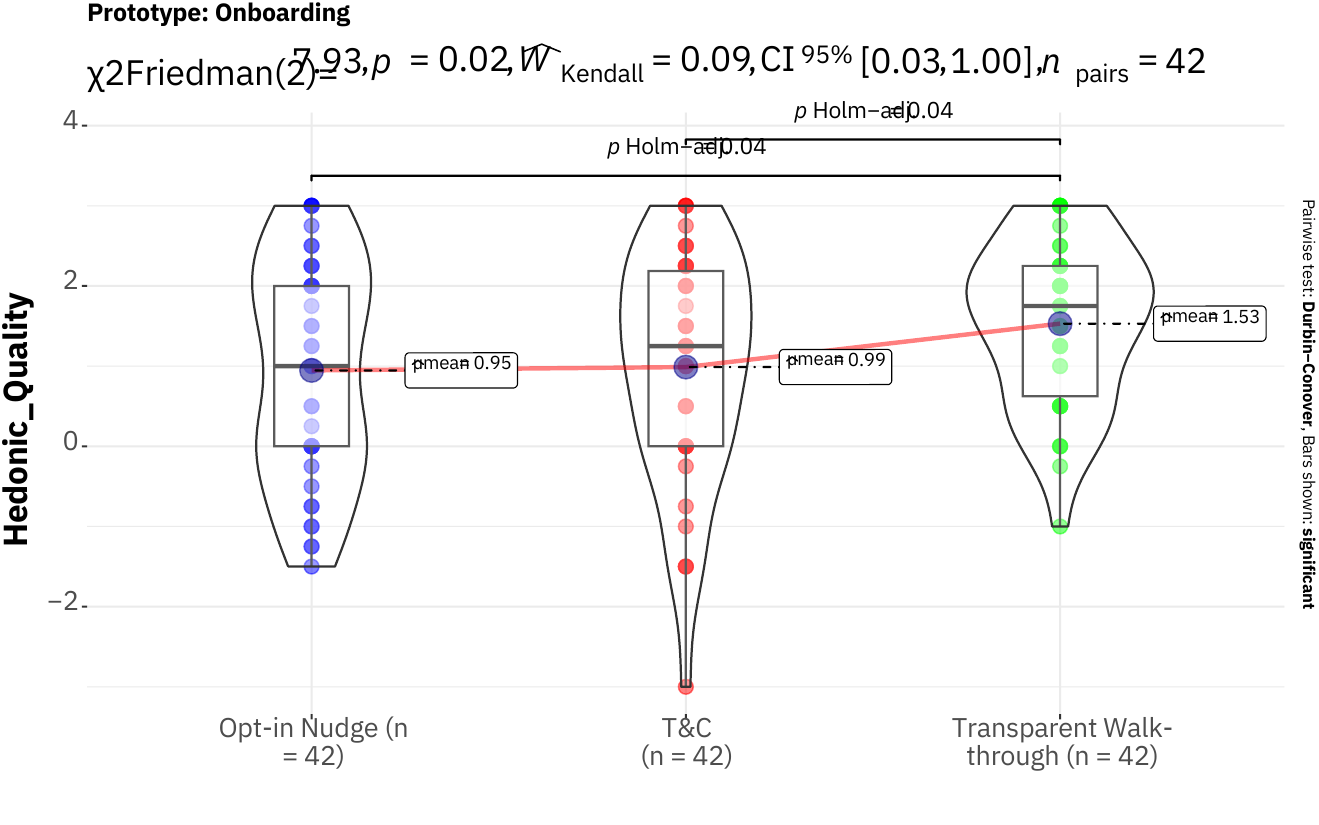}
\captionsetup{width=0.6\textwidth}   
\caption{User Experience Qualities of the Onboarding Prototypes.}
      \Description{User Experience Qualities of the Onboarding Prototypes.}
    \label{fig:OnboardingTrust}
\end{figure*}

\section{Discussion}
Our study provides insights into the role of transparent design in shaping user perceptions of privacy and trust in the context of in-car camera-based health monitoring systems. The findings suggest that transparency in design can help alleviate negative concerns. Specifically, our results indicate that a transparently designed onboarding procedure with clear and concise terms and conditions can significantly enhance user experience and trust in data use. This aligns with the existing literature on the inadequacies of ‘notice and consent’, and the need for more user-friendly and accessible privacy policies \cite{Bruening2015-rw}. This study demonstrates that policy design significantly impacts user experience design and should not be solely entrusted to lawyers, emphasizing the need for collaboration between law and design~\cite{Doherty2021-gz}.

The concept of ``privacy by design'' is essential in this context. By integrating privacy considerations into the design and architecture of IT systems and business practices, organizations can ensure that privacy is a core value from the outset \cite{Cavoukian2020-ud}. This approach helps build trust, meet regulatory requirements, and avoid potential legal issues.

Furthermore, the findings suggest that addressing feelings of creepiness and enhancing user control over personal data are crucial for user acceptance. This can be achieved by providing clear, understandable information about data collection and usage, as well as offering users meaningful choices regarding their data. 

The results also indicate that transparency can help alleviate users' concerns when the data use is legitimate, and the camera appears not to be the issue per se. Users' acceptance is more likely when they are assured of the legitimacy and necessity of data collection, and when they have control over their personal information. The best practice presented here can serve as a pattern for transparent user interface design~\cite{Sandhaus2023-so}.

\section{Limitations and Future Work}
Our findings suggest that transparency in design plays a crucial role in user acceptance of camera-based in-car health monitoring systems. Future research should aim for a more systematic exploration of design options by altering individual dimensions in a controlled manner. Focusing on individual dimensions such as transparency level, choice level, and deception level will allow for a more precise understanding of their individual impacts. 
Longitudinal studies could offer insights into how user perceptions and acceptance evolve over time with continued use of the technology. Exploring the impact of different cultural contexts on user acceptance is valuable, as privacy concerns and trust in technology can vary significantly across cultures. Additionally, integrating qualitative methods such as interviews or focus groups could provide deeper insights into user attitudes and concerns, informing the development of more effective design strategies. As the focus of this work was on the user interface design of monitoring feedback and onboarding, future work should also consider legal design and guidelines, including prototyping of information flows, policies, and policy text~\cite{Hagan2021-xu}.

\section{Conclusion}
In this study, we investigated the impact of transparent design on user perceptions in the context of in-car camera-based health monitoring systems. Our findings suggest that transparent design can significantly enhance user perceptions of privacy and trust and improve overall user experience. By providing clear and understandable information about data collection and usage, and offering users meaningful choices regarding their data, designers can help alleviate privacy concerns and build trust.

As in-car health monitoring systems become more prevalent, it is crucial to ensure that they are designed with user privacy and trust from the outset. Our findings will inform future research and practice, contributing to the development of more user-friendly and privacy-respecting data-driven technologies.

\begin{acks}
We express our gratitude to \href{https://www.meilitechnologies.com/}{Meili Technologies} for providing a reference for an In-Car Health Monitoring Platform and funding the participant incentives. Additionally, we extend our thanks to Prof. David Mimno for allowing us the flexibility to conduct this study in class.
\end{acks}

\bibliographystyle{ACM-Reference-Format}
\bibliography{manual}
  



\end{document}